\documentclass[12pt]{iopart}

\usepackage[colorlinks=true, linkcolor=blue]{hyperref}

\usepackage{iopams}

\expandafter\let\csname equation*\endcsname\relax

\expandafter\let\csname endequation*\endcsname\relax

\usepackage{amsmath}

\usepackage{graphicx}
\usepackage{bm}

\begin{document}

\title{Trade-off in perpendicular electric field control using negatively biased emissive end-electrodes}

\author{B. Trotabas and R. Gueroult}
\address{LAPLACE, Universit\'{e} de Toulouse, CNRS, INPT, UPS, 31062 Toulouse, France}
\ead{baptiste.trotabas@laplace.univ-tlse.fr}
\vspace{10pt}

\begin{abstract}
The benefits of thermionic emission from negatively biased electrodes for perpendicular electric field control in a magnetized plasma are examined through its combined effects on the sheath and on the plasma potential variation along magnetic field lines. By increasing the radial current flowing through the plasma thermionic emission is confirmed to improve control over the plasma potential at the sheath edge compared to the case of a cold electrode. Conversely, thermionic emission is shown to be responsible for an increase of the plasma potential drop along magnetic field lines in the quasi-neutral plasma. These results suggest that there exists a trade-off between electric field longitudinal uniformity and amplitude when using negatively biased emissive electrodes to control the perpendicular electric field in a magnetized plasma. 
\end{abstract}

%
%
%
%
%

\section{\label{sec:I} Introduction}

Electric fields perpendicular to magnetic surfaces in plasmas are of importance to a variety of applications. In magnetic confinement fusion for instance, radial electric fields are known to play an important role on confinement~\cite{Itoh1996}, as observed both in tokamaks~\cite{Taylor1989,Groebner1990,Oost2007} and stellerators~\cite{Stroth2001}. Confinement enhancement is in this case believed to be enabled by transport barriers induced by $\mathbf{E}\times\mathbf{B}$ sheared flows~\cite{Burrell1997,Terry2000,Burrell2020} in the presence of radial electric fields. Perpendicular electric fields also offer opportunities for the design of alternative confinement schemes in toroidal geometry such as the magnetoelectric confinement studied by Stix~\cite{Stix1971}, or more recently the wave driven rotating torus~\cite{Rax2017,Ochs2017}. In addition and beyond fusion, controlling perpendicular electric fields is essential for a growing number of applications of $\mathbf{E}\times\mathbf{B}$ configurations~\cite{Kaganovich2020}, and notably for the development of high-throughput plasma separation technologies~\cite{Gueroult2019,Zweben2018,Dolgolenko2017}. 

Although waves hold promise to produce such perpendicular electric fields~\cite{Fisch1992}, most of the experimental effort towards perpendicular electric field control to date has relied on electrode biasing. In magnetic confinement fusion experiments, the high plasma temperature and density generally prohibit inserting electrodes in the plasma core, and biasing experiments thus typically involve edge biasing~\cite{Weynants1993}. While this technique has been shown to be effective at affecting edge properties under certain conditions~\cite{Bagryansky2003,Schaffner2012}, the use of a single polarized surface at the edge - a biasing configuration known as a limiter -  does not provide control over how the applied bias distributes itself across magnetic surfaces. The perpendicular electric field indeed remains an intricate function of the plasma properties~\cite{Weynants1993}. Cooler and less dense plasmas, especially in open-field line geometries, open additional possibilities for biasing studies, and a broad array of electrode geometries have been used for the primary purpose of instabilities and turbulence suppression~\cite{Mase1991, Yamada2014,Desjardins2016} and flow control~\cite{Amatucci1996,Shinohara2007,Collins2012,Plihon2015,Gueroult2016,Desangles2021}. For perpendicular electric field control, a biasing configuration of particular interest is end-electrodes, that is electrodes intercepting magnetic field surfaces. The basic idea here, as originally suggested by Lehnert~\cite{Lehnert1970,Lehnert1973}, is that one could control the electric potential of individual magnetic field surfaces through the biases imposed on a set of end-electrodes. More specifically, the potential of a given magnetic surface is expected to be set by the applied bias on the electrode on which this magnetic surface terminates, allowing in principle in turn for perpendicular electric field control. While very attractive, the practicality of this scheme remains a question. Indeed, while control has been successfully demonstrated under certain conditions~\cite{Tsushima1986,Bardakov2014}, other experiments reported more contrasted results (see Ref.~\cite{Gueroult2019} for a more extensive discussion of end-electrodes biasing experiments in linear geometry). 

Conceptually, the ability to control the potential of a magnetic surface in a plasma through the bias applied on an end-electrode can be split into two problems: controlling the potential drop along magnetic field lines in the quasi-neutral plasma and controlling the potential drop across the non-neutral sheath formed in front of the biased electrode. So far, these two problems have mostly been treated separately. On the former, control along magnetic field lines in a quasi-neutral plasma is guaranteed in the limit that field lines are isopotential, which as noted by Lehnert in his original paper~\cite{Lehnert1970} can in principle be asymptotically approached using a large enough magnetic field. Restated through conductivities, this is equivalent to the limit of a zero perpendicular to parallel conductivity ratio $\mu=\sigma_{\perp}/\sigma_{\parallel}$. Practically though, $\mu$ is finite, which implies that field lines are not strictly isopotential. Examining specifically this problem, it has recently been shown that the relative variation in potential along field lines in a plasma column of radius $a$ and length $L$ is about $\tau=L/a\sqrt{\mu}$~\cite{Gueroult2019a}.  This suggests that a necessary condition for potential control along field lines is $\tau\ll 1$. On the latter, Liziakin \emph{et al.} showed considering the sheath formed in front of a negatively biased electrode that the combination of the plasma perpendicular conductivity $\sigma_{\perp}$ and the ion saturation current sets a lower limit on the minimal plasma potential $\phi_p<0$ one can expect from applying a negative bias $\phi_e<\phi_p$~\cite{Liziakin2020}. Building on this finding, the same authors recently showed that the addition of thermionic emission from a negatively biased electrode can lower further the plasma potential $\phi_p$ at the expense of the voltage drop across the sheath $\phi_p-\phi_e$~\cite{Liziakin2021}, which is consistent with earlier theoretical work~\cite{Poulos2019} and observations~\cite{Jin2019}.

In this paper, we consider these two problems in a unified model with the goal of highlighting the overall limits on perpendicular electric field control from negatively-biased end-electrodes, and examine in particular the effect of thermionic emission. In Section~\ref{sec:II}, we briefly introduce our model. In Section~\ref{sec:III}, the sheath models developed by Liziakin \emph{et al.}~\cite{Liziakin2020,Liziakin2021} are first used to identify what the limits on potential control at the sheath edge are. In Section~\ref{sec:IV}, these insights from sheath dynamics are then coupled into models for the potential distribution in a quasi-neutral magnetized plasma to highlight the influence of the sheath, and theoretical predictions are compared to numerical simulations. In Section~\ref{sec:V}, the main results are summarized.

\section{\label{sec:II} Model description}

\begin{figure}[htbp]
    \centering
    \includegraphics[scale = 0.90]{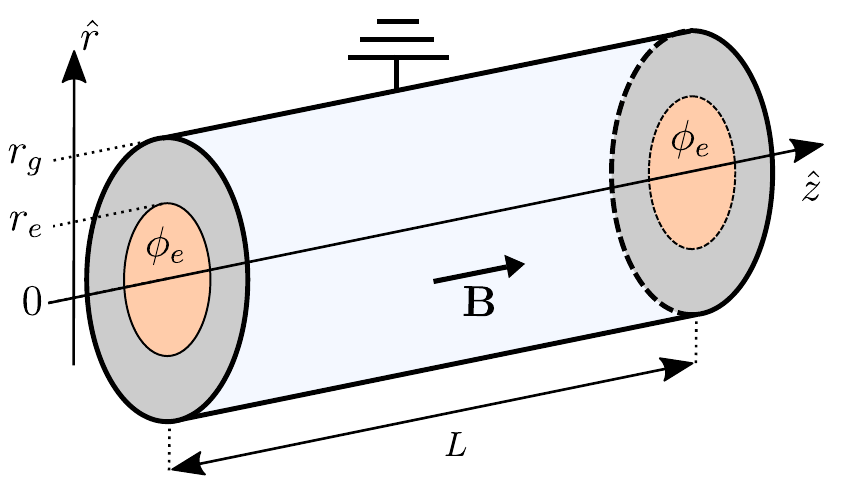}
    \caption{Magnetized plasma column terminating on two single disk electrodes (in orange). Electrodes are biased negatively (potential $\phi_e$) with respect to the grounded vacuum vessel. The entire domain is permeated by a uniform axial magnetic field $\mathbf{B} = B_0 \bm{\hat{z}}$.}
    \label{fig:Scheme_Plasma_colum}
\end{figure}

The configuration studied in this work is illustrated in figure~\ref{fig:Scheme_Plasma_colum}. It consists in a symmetrical linear machine with single full disks electrodes (sometimes referred to as button electrodes) terminating axially a magnetized plasma column. We note $L$ the inter-electrode distance and $r_e$ and $r_g$ the radii of the biased electrode and grounded vacuum vessel, respectively. We further limit ourselves in this work to negative biases imposed on the disk electrodes, but consider both cold and hot surfaces to highlight the effect of thermionic emission. This simpler biasing configuration is used here to underline some of the key features expected in the more complex segmented concentric ring electrodes configuration proposed by Lehnert~\cite{Lehnert1970}.

The plasma filling this volume is assumed to be produced by an external source (\emph{e~.g.} radio-frequency or ECR) which is not modeled in this work, and we further assume that the biased electrodes do not affect the plasma properties other than through the plasma potential. In addition, we consider the plasma to be uniform and quiescent, that is that density and temperature gradients, as well as possible instabilities, are neglected. Away from the sheaths formed in front of the biased electrodes, the plasma is therefore modeled as a uniform anisotropic media characterized by parallel and perpendicular conductivities $\sigma_{\parallel}$ and $\sigma_{\perp}$. Finally, since our system is symmetrical, only the left-hand side of the domain is studied, that is $\left [0, r_g \right ] \times \left [ -L/2, 0 \right ]$. 

\begin{figure}[htbp]
    \centering
    \includegraphics[scale = 1]{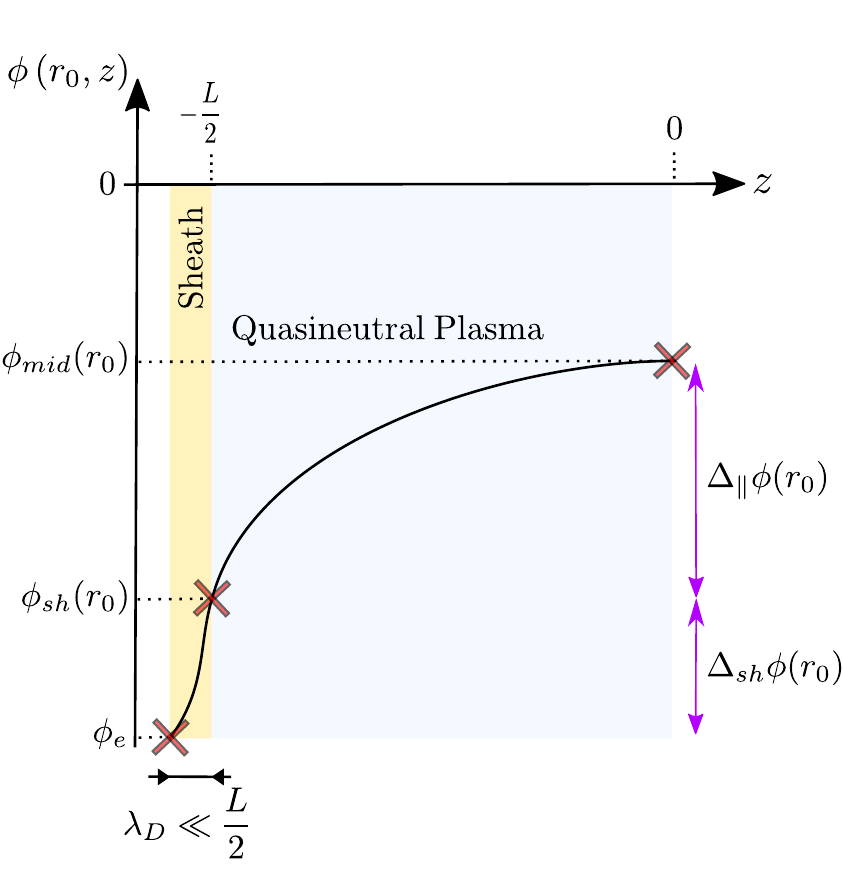}
    \caption{Sketch of the variation of the electric potential $\phi$ along a magnetic field line (radius $r_0<r_e$) between the negatively biased electrode and the midplane. The total voltage drop $\phi_{mid}(r_0)-\phi_e$ is the sum of a voltage drop across the sheath $\Delta_{sh}\phi(r_0)$ and a voltage drop along field line in the quasi-neutral plasma $\Delta_{\parallel}\phi(r_0)$, from Eqs.~(\ref{Eq:Delta_sh}) and (\ref{Eq:Delta_parra}). }
    \label{fig:Scheme_Plasma_colum_part2}
\end{figure}

Having perpendicular electric field control in mind, a key element of the plasma potential response as a function of the applied bias $\phi_e$ and the plasma parameters (through the conductivities) is the evolution of potential along magnetic field lines. For the uniform magnetic field considered here this means the variation of $\phi(r,z)$ at constant radius $r_0$ as illustrated in figure~\ref{fig:Scheme_Plasma_colum_part2}. To facilitate interpretation, we introduce the short-hands $\phi_{sh}(r) = \phi(r,-L/2)$ and $\phi_{mid}(r) = \phi(r,0)$ for the plasma potential radial profile at the sheath edge and in the mid-plane, respectively. With this notation the voltage drop across the sheath and along field lines in the quasi-neutral plasma are then simply
\begin{equation}
    \Delta_{sh} \phi (r) =  \phi_{sh} \left (r\right ) - \phi_e
    \label{Eq:Delta_sh}
\end{equation}
and
\begin{equation}
    \Delta_\parallel \phi(r) = \phi_{mid} \left ( r\right ) - \phi_{sh} \left ( r\right ).
    \label{Eq:Delta_parra}
\end{equation}
Finally, in an effort to ease comparison between experiments across a broad range of conditions, we work in this study with dimensionless variables and define for all electric potential quantities $\phi$ their dimensionless analog $\psi$ through $\psi= \phi/T_e$ with $T_e$ in eV. 

\section{\label{sec:III} Effect of the sheath on the plasma potential at the sheath edge}

Inserting a biased electrode in contact with a plasma leads to the formation of a sheath, that is to a non-neutral region a few Debye length thick connecting the bulk plasma to the electrode~\cite{baalrud2020interaction} (see figure~\ref{fig:Scheme_Plasma_colum_part2}). For biased electrodes of sufficiently small surface area, the bulk plasma parameters, and in particular the plasma potential, are to zero-th order unaffected by the electrode bias. This is an essential hypothesis in probe theory~\cite{Hershkowitz1989}. On the other hand, for large enough surface areas, the bulk plasma potential can be affected by the applied bias~\cite{baalrud2020interaction}. The thin sheath region is then expected to control to which extent the applied bias potential is passed along field lines into the plasma bulk.

\subsection{\label{sec:III_A} Sheath transfer function for the potential: saturated and non-saturated regimes}

A first picture of how the plasma potential at the sheath edge relates to the applied bias can be obtained from the model proposed by Liziakin \emph{et al.}~\cite{Liziakin2020}. In this model the potential $\phi(r,z)$ in the plasma column shown in figure~\ref{fig:Scheme_Plasma_colum} is assumed to be independent of $z$, which corresponds to the limit $\mu = \sigma_{\perp}/\sigma_{\parallel}\rightarrow 0$. In this case one simply gets $\phi(r,z) = \phi_{sh}(r)$. This model further assumes that the potential is constant in the shadow of the electrode (that is for $r\leq r_e$), allowing to model the discharge as a constant radial current $I$ flowing from $r_e$ to $r_g$ across magnetic surfaces. The influence of this last simplifying hypothesis is examined in detail via a more complete model in Section~\ref{sec:III_C}.

\begin{figure}[t]
    \centering
    \includegraphics[scale = 1]{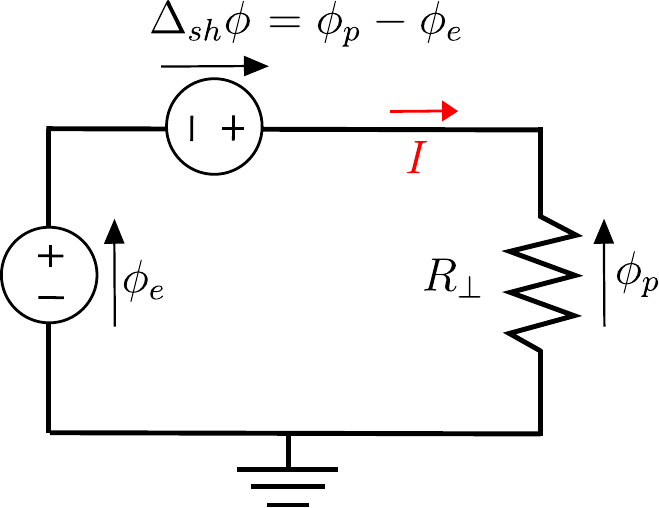}
    \caption{Equivalent electric circuit for the plasma potential $\phi_p$ in the shadow of the biased electrode for an axially uniform plasma proposed by Liziakin \emph{et al.}~\cite{Liziakin2020}. }
    \label{fig:0D equivalent electric circuit}
\end{figure}

Under these assumptions, the plasma potential in the electrode's shadow $\phi_p = \phi(r\leq r_e,z)$ is simply $R_{\perp}I$, where the perpendicular resistance $R_{\perp}$ opposing the current has been obtained by integrating between $r_e$ and $r_g$ the incremental resistance
\begin{equation}
    dR(r) = \frac{dr}{\pi L \sigma_\perp r}
    \label{dR}
\end{equation}
associated with the annular region of length $L$ located between $r$ and $r+dr$~\cite{Liziakin2020}. Note here that $R_{\perp}$ is the perpendicular resistance associated with half the plasma column length, consistent with the fact that we consider $I$ as the current on one axial end-electrode. This plasma potential $\phi_p$ minus the voltage drop across the sheath $\Delta_{sh}\phi$ must be equal to the applied bias $\phi_e$, as illustrated in figure~\ref{fig:0D equivalent electric circuit}. In addition, current continuity requires for this current $I$ to be equal to the current drawn at the biased electrode. Considering here an ion sheath formed in front of a hot negatively biased electrode and counting positively particle fluxes leaving the electrode, one gets $I = I_e - I_{is} - I_{eth}$ where electron, ion and thermionic electron currents have respectively been obtained from the surface integrated current densities 
\begin{subequations}
\begin{eqnarray}
j_e & = & \displaystyle j_{is} \exp \left ( \Lambda + \frac{\phi_e-\phi_p }{T_e} \right ) , \label{je}
\\
j_{is} & = & \displaystyle \frac{e n c_s}{2}, \label{jis}
\\
j_{eth} & = & A_G {T_W}^2 \exp\left( -\frac{eW}{k_B T_W} \right ). \label{jeth}
\end{eqnarray}
\end{subequations}
Here $n$ is the plasma density, $c_s = \sqrt{e T_e / m_i}$ is the ion sound speed, $\Lambda = \ln ( \sqrt{2 m_i / ( \pi m_e )})$ is a sheath parameter, $T_W$ is the temperature of the electrode, $W$ is the work function of the material and $A_G$ is the Richardson's constant taken here equal to $6.0\times 10^5 \mathrm{A\cdot K^{-2}\cdot m^{-2}}$~\cite{Richardson1912}. Moving to dimensionless variables, the normalized current flowing through the plasma hence writes
\begin{equation}
    \frac{I}{I_{is}} =  \exp \left ( \Lambda + \psi_e - \psi_p \right ) - 1 - \Xi 
    \label{I_electrode}
\end{equation}
with $\Xi = j_{eth}/j_{is}$ a dimensionless parameter quantifying thermionic emission with respect to the ion saturation current. For an applied bias $\psi_e\leq\psi_p$ the current $I$ is negative and reaches a minimum equal to $-\left ( I_{is} + I_{eth} \right )$ for sufficiently large $\psi_p-\psi_e$. Note that we chose here to ignore for simplicity possible dependencies of $j_{eth}$ on $\phi_p$ through the Schottky effect~\cite{Nottingham1956}, as well as possible dependencies of $j_{is}$ on $\phi_p$ through a modification of the Bohm velocity at the sheath boundary in the presence of thermionic emission~\cite{palacio2013bohm}. 

Plugging Eq.~(\ref{I_electrode}) into Ohm's law across the plasma resistance $\phi_p = R_{\perp} I$ and moving to dimensionless variables finally yields a transcendental equation for the normalized plasma potential
\begin{equation}
    \exp \left ( \Lambda + \psi_e - \psi_p \right ) - 1 - \chi \psi_p - \Xi= 0 \text{ ,}
    \label{Vp_adim}
\end{equation}
where we have defined $\chi = T_e \big / I_{is}R_\perp$. Eq.~(\ref{Vp_adim}) is identical to that obtained by Liziakin \emph{et al.}~\cite{Liziakin2020}, other than for the choice of using dimensionless variables and the addition in this work of thermionic emission.

\begin{figure}[htbp]
    \centering
    \includegraphics[scale = 1]{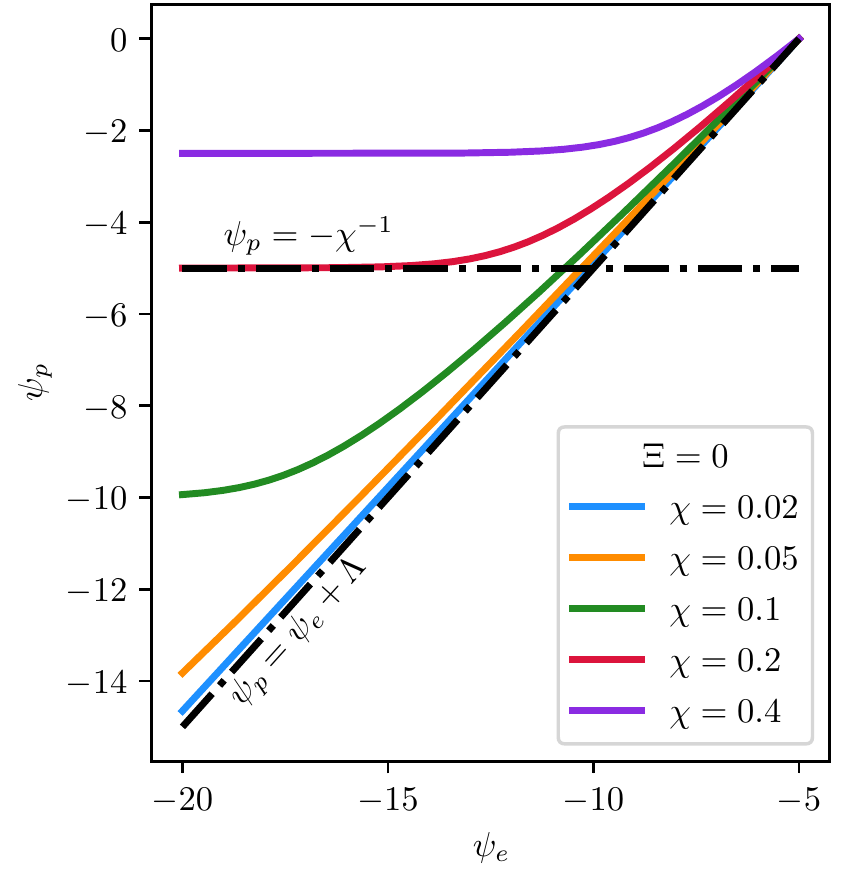}
    \caption{Normalized plasma potential in the electrode shadow $\psi_p$ as a function of the normalized applied bias $\psi_e$ for different values of $\chi$ as predicted in Eq.~(\ref{Vp_adim}) absent thermionic emission ($\Xi=0$). Potentials are normalized by the electron temperature $T_e$ in eV, $\psi=\phi/T_e$. }
    \label{fig:saturation_criteria_NoneXi}
\end{figure}

Considering first the case without thermionic emission ($\Xi=0$), figure~\ref{fig:saturation_criteria_NoneXi} shows the evolution of $\psi_p$ as a function of $\psi_e$ for different values of $\chi$ as predicted by Eq.~(\ref{Vp_adim}). For small biases applied at the electrode $\psi_e$, or a sufficiently small value of $\chi$, the plasma potential is seen to follow the applied bias, that is $\psi_p = \psi_e + \Lambda$. In this case the plasma potential is hence controlled by the electrode bias. The voltage drop across the sheath is then small, $\Delta_{sh} \psi = \Lambda$, and the current drawn at the electrode is negligible compared to $I_{is}$. For reasons that will become clear in the next paragraph, we refer to this regime as the non-saturated regime.

For larger negative biases for a given value of $\chi$, figure~\ref{fig:saturation_criteria_NoneXi} shows that the plasma potential progressively deviates from the floating solution $\psi_p = \psi_e + \Lambda$ until it reaches a minimal value $\psi_p = \chi^{-1}$. This behaviour is consistent with the fact that the current $I<0$ then approaches its minimum value $-I_{is}$. At this point the plasma potential $\psi_p$ is no longer controlled by the applied bias $\psi_e$, and any further decrease in applied bias $\psi_e$ is entirely recovered in the voltage drop across the sheath $\Delta_{sh}\psi$ which grows as $|\psi_e|-\chi^{-1}$. Reflecting the property of the current in this regime, we refer to it as saturated or current-limited regime. In contrast with the non-saturated regime, the minimum plasma potential is indeed here set by the maximum current drawn at the biased electrode. The dimensionless parameter $\chi^{-1}$ can then be interpreted as a measure of the maximum radial voltage drop the plasma column can sustain. From the definition of $R_{\perp}$, one gets 
\begin{equation}
    \chi = \frac{L}{r_e^2  \ln ( r_g / r_e )} \frac{T_e  \sigma_\perp}{j_{is}}.
    \label{chi}
\end{equation}

The transition from non-saturated to saturated regime, that is from a plasma potential controlled by the electrode bias to a plasma potential controlled by plasma parameters, takes place for $|\psi_e|\chi\geq 1$. The boundaries in physical operating parameters space (densities, temperatures, bias) for this regime transition can be obtained by developing the parametric dependencies of $j_{is}$ and $\sigma_{\perp}$ for the plasma conditions under consideration, as it will be done in Section~\ref{sec:IV}.

\subsection{\label{sec:III_B} Additional control offered by thermionic emission}

Examining now specifically the effect of thermionic emission, Eq.~(\ref{Vp_adim}) shows that the plasma potential in the saturated regime, that is the minimal plasma potential no matter how negative the applied bias, writes
\begin{equation}
    \psi_p^{sat} = -\frac{1 + \Xi}{\chi}.
    \label{psi_sat_regime}
\end{equation}
Compared with the case absent thermionic emission studied above, one finds that its amplitude is $(1+\Xi)$ times larger. This result is simply the consequence that the discharge current is $1+\Xi$ larger when taking into consideration thermionic emission, while the perpendicular plasma resistance $R_{\perp}$ remains the same.

\begin{figure}[htbp]
    \centering
    \includegraphics[scale = 1]{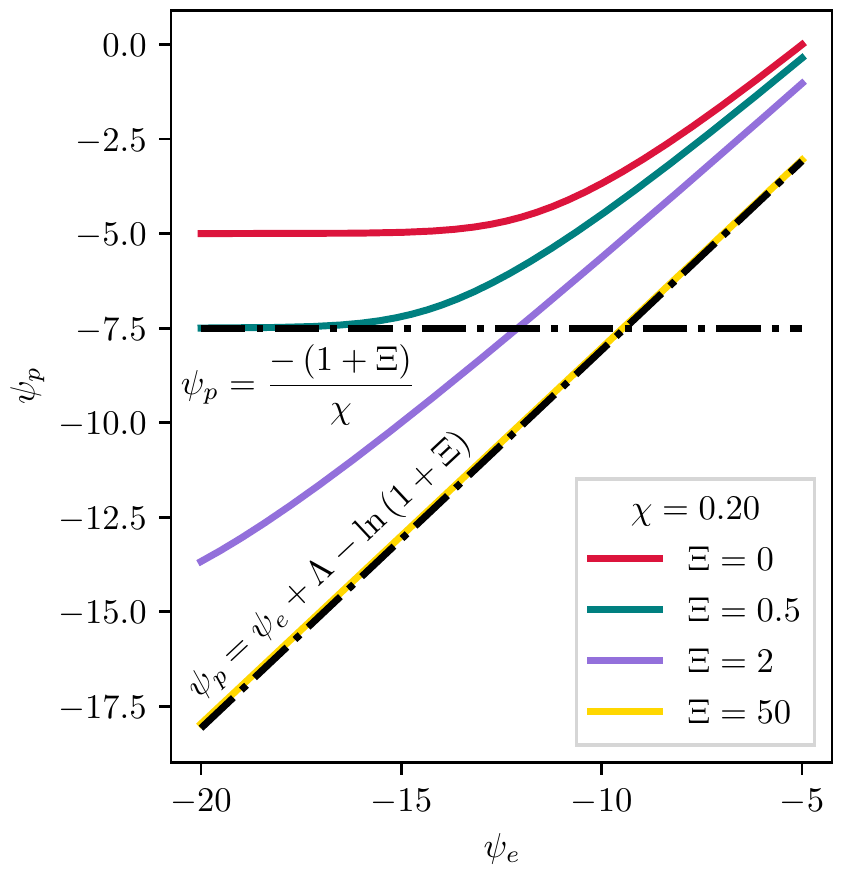}
    \caption{Normalized plasma potential in the electrode shadow $\psi_p$ as a function of the normalized applied bias $\psi_e$ for different values of thermionic emission $\Xi$, as predicted in Eq.~(\ref{Vp_adim}). Potentials are normalized by the electron temperature $T_e$ in eV, $\psi=\phi/T_e$.}
    \label{fig:saturation_criteria_Xi}
\end{figure}

Since $\psi_p^{sat}$ by definition coincides with the applied bias $\psi_e$ below which the transition from non-saturated to saturated regime occurs, the above result suggests a larger range of accessible plasma potential for a given $\chi$ (that is for a given set of plasma parameters). Put differently, as illustrated in figure~\ref{fig:saturation_criteria_Xi}, increasing thermionic emission for a given bias and a given $\chi$ allows transitioning from a saturated regime to a non-saturated regime, and therefore regaining control over the plasma potential. In the process the plasma potential decreases and progressively approaches the applied bias as thermionic emission increases, consistent with experimental observations~\cite{Jin2019}.

Note that while the added control offered by thermionic emission suggests operating with large thermionic currents $I_{eth}$, there is practically a limit to how large this current can be. Indeed, past a certain value a virtual cathode is expected to form in front of the electrode. This will in turn naturally limit the thermionic current reaching the bulk plasma, and thus the thermionic contribution to the discharge current. The value of $I_{eth}$ (and thus of $\Xi$) for which this virtual cathode is expected to form can be derived analytically by solving Poisson's equation in the sheath region~\cite{Poulos2019}.

\subsection{\label{sec:III_C} Radial potential profile}

Up to this point we have only examined the relation between the applied bias $\phi_e$ and the plasma potential in the electrode shadow $\phi_p$. This is because in the model introduced in Section~\ref{sec:III_A} the potential is assumed constant for $r\leq r_e$ and the current constant for $r>r_e$. Integration of Ohm's law with the incremental resistance given in Eq.~(\ref{dR}) hence leads to a simple logarithmic profile for $\phi_{sh}$ connecting $\phi_p$ for $r<r_e$ to the ground at $r=r_g$. 

\begin{figure*}[htbp]
    \centering
    \includegraphics{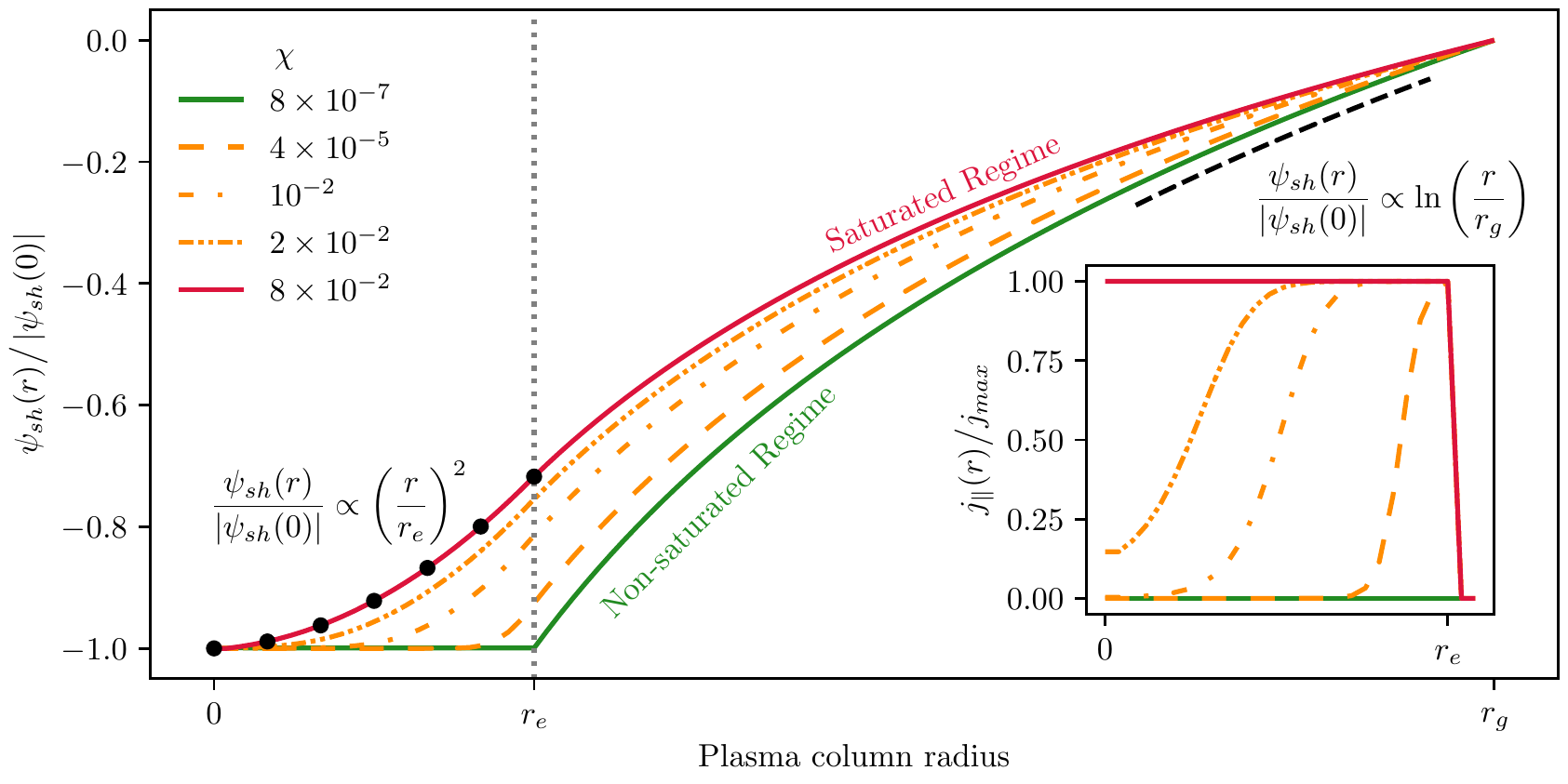}
    \caption{Normalized plasma potential $\psi_{sh}(r) \big /  |\psi_{sh}(0) |$ radial profile for different values of $\chi$, and normalized current density radial profile (inset). Green and red curves highlight the potential profiles obtained for the non-saturated and saturated regimes, respectively. The black dots for $r\leq r_e$ illustrate a parabolic radial dependence, whereas the dotted black curve near $r_g$ represents a logarithmic radial profile. }
    \label{fig:Radial_profile_psi_p}
\end{figure*}

A more physical model has recently been proposed by Liziakin \emph{et al.}~\cite{Liziakin2021} noting that the radial current $I(r)$ at any radius must be equal in steady state to the surface integrated current density for all radii $r'<r$, that is
\begin{equation}
    I(r) = \displaystyle \int_{0}^{r} j_{sh,\parallel}(r') 2 \pi r' dr'
    \label{Ir(r)}
\end{equation}
with
\begin{equation}
j_{sh,\parallel}(r) = j_e(r) - j_{is} - j_{eth} 
\label{j_para(r)}
\end{equation}
the current density through the sheath. By implicitly assuming that there is no current source past the outer radius of the electrode $r_e$, the requirement for a constant potential in the electrode shadow can then be lifted, and the plasma potential obtained by integrating from the ground reference at $r=r_g$ to any radius $r$ the local Ohm's law $d\phi_{sh}(r)=I(r)dR(r)$. Plugging in Eq.~(\ref{Ir(r)}) and going back to dimensionless variables yields a linear integro-differential equation for $\psi_{sh}$,
\begin{equation}
    \displaystyle \frac{d \psi_{sh}(r)}{dr}  =\int_0^r\mathcal{F}(r',\psi_{sh}(r'))dr'.
    \label{integro_differential_eq}
\end{equation}
Here $\mathcal{F}$ is a function that depends on the plasma parameters through $T_e$ and $\sigma_{\perp}$, on the plasma composition through the sheath parameter $\Lambda$, on the applied bias $\psi_e$, on the thermionic current parameter $\Xi$ and on the geometric parameters $r_e$, $r_g$ and $L$. In Ref.~\cite{Liziakin2021}, Eq.~(\ref{integro_differential_eq}) was solved numerically for spatially dependent plasma parameters (plasma density and electron temperature) measured in an experiment, as a way to validate potential profile predictions against experimental data. Here we similarly solve Eq.~(\ref{integro_differential_eq}) using a shooting method but instead use these results to highlight some characteristic properties of the radial potential obtained for a uniform plasma (\emph{i.~e.} uniform $\sigma_{\perp}$), as shown in figure~\ref{fig:Radial_profile_psi_p}. Another important distinction is that Liziakin \emph{et al.} focused specifically in Ref.~\cite{Liziakin2021} on the case where $\sigma_{\perp}$ depends linearly on $d\psi_{sh}(r)/dr$ by assuming that ions drift with velocity ${B_0}^{-1}d \phi_{sh}(r)/dr$ in a background static neutral gas. This corresponds to the limit of supersonic ion rotation $\Omega r\gg v_{thi}$. In contrast we purposely do not consider here for the moment a particular mechanism for cross-field conductivity, but we do assume that $\sigma_{\perp}$ does not depend on $\phi$ nor its radial derivatives. We simply note here that this hypothesis may be representative of either ion-neutral collisions in the slow rotation limit $\Omega r\ll v_{thi}$, or of a different conductivity mechanism (\emph{e.~g.} electron-neutral collisions as examined by Liziakin \emph{et al.} in an earlier study~\cite{Liziakin2020}).

Looking first at the normalized current density, one recovers the non-saturated and saturated regimes for respectively small and large values of $\chi$. Indeed, consistent with the simpler model used in Sections~\ref{sec:III_A} and \ref{sec:III_B}, the current density on the electrode $j_\parallel(r)$ is observed to be uniform and equal to its maximum value (in absolute value) in the saturated regime, whereas it is negligible in the non-saturated regime. However, we also observe that these two regimes are now separated by a group of curves obtained for intermediate values of $\chi$ for which the current goes from near zero on-axis to its saturated value at the outer edge of the biased electrode. This shows that saturation is actually a local phenomenon. By analogy with the nomenclature introduced earlier, we refer to this regime as partially saturated. A closer examination reveals that saturation first appears at the outer edge of the biased electrode and progressively moves radially inward with increasing $\chi$ until full saturation has been reached, which is to be expected for a monotonically increasing radial potential profile.

Moving on now to the normalized plasma radial profile and starting from small value of $\chi$, that is in the non-saturated regime, one observes a nearly constant profile in the shadow of the electrode ($r<r_e$). This result can be explained from Ohm's law and the fact that the current drawn at the electrode is in this case very small. In this regime in the electrode shadow one simply finds $\psi_p = \psi_e+\Lambda$. This same profile holds as $\chi$ increases up until partial saturation begins at the outer edge of the electrode. As the current then becomes larger (in amplitude), Ohm's law predicts an increase of $\psi_{sh}$. Since, as showed above, partial saturation move radially inward with $\chi$, the position where $\psi_{sh}(r)$ begins to deviate from its constant on-axis value is also observed to move radially inward. Finally, once the fully saturated regime has been reached, a new profile sets in and holds for arbitrary large value of $\chi$. 

Because the saturated regime is characterized by a saturated current at the electrode, the flux of electrons reaching the electrode is then by definition null. This property allows for further analysis. Indeed, Eq.~(\ref{integro_differential_eq}) then becomes a simple ordinary differential equation whose solution for $r\leq r_e$ writes
\begin{equation}
    \psi_{sh}^{sat}(r) =  -\frac{ 1 + \Xi}{\chi}\left[1+\frac{1}{2 \chi \ln \left ( r_g / r_e \right )} \left ( 1 - \frac{r^2}{r_e^2} \right )\right].
\label{psi_p_Liziakin2021}
\end{equation}
Here one recognizes that the first term in brackets on the right hand side is $\psi_{sh}^{sat}(r_e)$ given in Eq.~(\ref{psi_sat_regime}), which was assumed to be the constant potential found in the electrode shadow in Sections~\ref{sec:III_A} and \ref{sec:III_B}. In contrast it is found here that the potential in the electrode shadow actually exhibits a parabolic radial dependence. This is noteworthy insofar that this profile leads to a constant angular $\mathbf{E}\times\mathbf{B}$ drift frequency, and thus to solid body rotation in the shadow of the electrode assuming crossed-field drift is the dominant contribution to rotation. It should also be noted here that the fact that a parabolic profile is obtained in Eq.~(\ref{psi_p_Liziakin2021}) instead of the $r^{3/2}$ profile derived by Liziakin \emph{et al.}~\cite{Liziakin2021} is the consequence that, as mentioned above, $\sigma_{\perp}$ is assumed in this work independent of $d \phi_{sh}(r)/dr$.

Finally, one notes that all profiles feature a logarithmic radial dependence past the outer radius of the electrode $r_e$, no matter the regime. This is the direct consequence of the fact that we assumed zero axial current past $r_e$, so that the radial current $I(r)$ in Eq.~(\ref{Ir(r)}) is constant for $r>r_e$. In this case the potential simply writes
\begin{equation}
\psi_{sh}(r>r_e)  =  I(r_e) \displaystyle \int_{r_g}^{r} \frac{dr}{\pi L \sigma_\perp r}
\end{equation}
which indeed yields the observed logarithmic radial dependence, consistent with Ref.~\cite{Liziakin2020}. Profiles then only differ through their value at $r=r_e$, which is a function of the total current emitted by the electrode $I(r_e)$. 

To summarize our findings, we have shown in this section that the ability to pass the negative applied bias $\psi_e$ unchanged (up to the sheath parameter $\Lambda$) along field lines into the plasma bulk is conditioned upon the smallness of the quantity $|\psi_e|\chi(1+\Xi)^{-1}$. The value of this quantity can be obtained for actual plasma parameters through Eq.(\ref{chi}) and the definition of $\Xi$. For values of $\psi_e$ such that this quantity is greater than $1$, the regime is said to be saturated. The plasma potential then no longer varies with $\psi_e$, $\psi_p = \psi_p^{sat}$, whereas the sheath drop $\Delta_{sh} = \psi_p-\psi_e$ grows as $|\psi_e|-(1+\Xi)\chi^{-1}$. These results point to the value of thermionic emission to increase potential control. Finally, a more refined model of currents in the system shows that the saturation phenomenon is local, starting at the outer edge of the biased electrode and moving radially inward until the entire surface of the biased electrode is collecting maximum current density.

\section{\label{sec:IV} Effect of the sheath on the voltage drop along field lines}

Controlling the potential drop across the sheath is essential for perpendicular electric field control through end-electrodes biasing but it is not enough. One indeed must also ensure that the potential does not vary significantly along field lines in the quasi-neutral plasma. Assuming a monotonic variation of $\phi(r,z)$ along $z$, this condition translates into the smallness of $\Delta_{\parallel}\phi(r) = \phi_{mid}(r)-\phi_{sh}(r)$. Considering only the quasi-neutral plasma (\emph{i.~e.} neglecting sheath effects), Gueroult \emph{et al.}~\cite{Gueroult2019} showed that the normalized voltage drop along field lines, that is $\Delta_{\parallel}\phi(r)/\phi_{sh}(r)$, is small under the condition $\tau = L/r_g\sqrt{\sigma_\perp/\sigma_\parallel}\ll 1$. This condition, however, does not provide a measure of $\Delta_{\parallel}\phi(r)$ since it depends on $\phi_{sh}(r)$ which is governed by the sheath. In this section we revisit this question in light of the insights in sheath physics obtained in Section~.\ref{sec:III}. 

\subsection{\label{sec:IV_A} Insights from theory}

Ohm's law at the sheath writes
\begin{equation}
    \displaystyle \left.\begin{matrix} \displaystyle j_\parallel \left (r, - \frac{L}{2} \right ) = \displaystyle - \sigma_\parallel \frac{\partial \phi}{\partial z} \end{matrix}\right|_{z = -L/2}.
\end{equation}
Although local, this results suggests that the voltage drop along field lines in the quasi-neutral plasma will depend on the current collected at the electrode. In fact, one shows as done in \ref{app:A} that in the particular case that the current density through the sheath does not depend on the radius, which is precisely verified in the saturated regime, this result can be generalized to obtain to lowest order in $\tau$ the global result
\begin{equation}
    \Delta_\parallel \psi^{sat} = \frac{L}{4 T_e \sigma_\parallel} j_{is}   \left (1 + \Xi \right ).
    \label{delta_para_psi_Ohm}
\end{equation}
This result shows that, at least in the saturated regime, the voltage drop along field lines in the quasi-neutral plasma $\Delta_\parallel \psi$ is expected to grow with thermionic emission. 

Taking a step back, we have seen in Section~\ref{sec:III} that thermionic emission can help achieve control over the plasma potential over a larger range of operating conditions compared to the case of a cold biased electrode. In has notably been shown to help limit the voltage drop which exists across the sheath in the case of a strongly negative bias such that $|\psi_e|\chi>1$. On the other hand, Eq.~(\ref{delta_para_psi_Ohm}) now suggests that in these same saturated conditions thermionic emission would lead to a greater voltage drop along field lines. To better understand this apparent trade-off and to which extent thermionic emission is desirable for perpendicular electric field control, we now turn to numerical simulations.

\subsection{\label{sec:IV_B} Numerical simulations}

\begin{figure}[htbp]
    \centering
    \includegraphics[scale = 1]{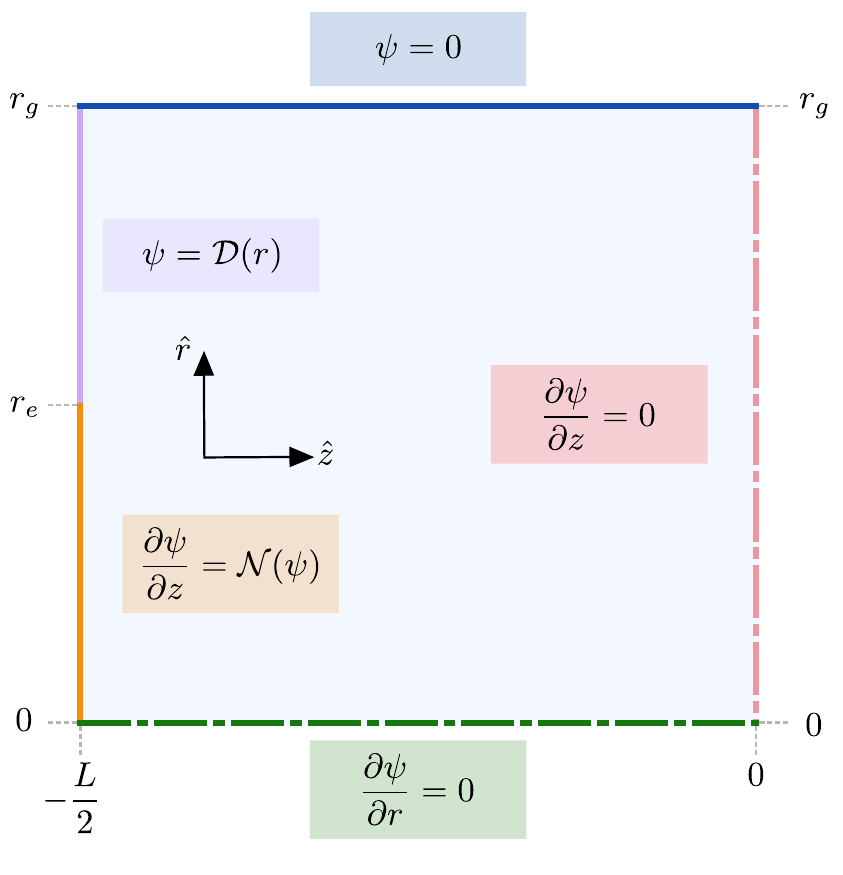}
    \caption{Computational domain used for numerical simulations. The green dash-dot boundary condition (BC) corresponds to axisymmetry. The pink dash-dot BC corresponds to symmetry with respect to the plane $z=0$. The blue BC represents the grounded vacuum vessel. The orange BC corresponds to the flux condition Eq.~(\ref{Neumann}) modeling the ion-sheath in front of the electrode. The purple BC is the Dirichlet condition given in Eq.~(\ref{Dirichlet}). }
    \label{fig:FDM_BC}
\end{figure}

While the anisotropic Laplace equation
\begin{equation}
    \frac{1}{r}\frac{\partial}{\partial r}\left[r\frac{\partial \phi}{\partial r}(r,z)\right] + \frac{\sigma_{\parallel}}{\sigma_{\perp}}\frac{\partial^2 \phi}{\partial z^2}(r,z) = 0
\label{Eq:Laplace_anis}
\end{equation}
obtained from the combination of Ohm's law for a static background neutral $\bm{j}=\bm{\sigma}\mathbf{E}$ and charged conservation $\bm{\nabla}\cdot\bm{j}=0$ allowed for analytical solutions when imposing Dirichlet conditions~\cite{Gueroult2019a}, the use of more physical flux conditions requires numerical modeling. Eq.~(\ref{Eq:Laplace_anis}) is thus solved here using finite differences in the interior of the domain shown in figure~\ref{fig:FDM_BC} and implementing flux conditions at the electrodes in a way very similar to that employed by Von Compernolle \emph{et al.}~\cite{van2019modifications}. Specifically, noting $z_0 = -L/2$ the axial position of the left boundary of the domain, the ion-sheath in front of the electrode is modeled via the non-linear Neumann condition
\begin{equation}
 \left.\frac{\partial \psi(r,z)}{\partial z}\right|_{z=z_0}= \frac{j_{is}}{T_e \sigma_\parallel} \left [ 1 + \Xi- \exp \left ( \Lambda + \psi_e(r) - \psi (r,z_0)\right ) \right ].
\label{Neumann}
\end{equation}
In addition, we model the axial boundary in the electrode plane $z=z_0$ between $r_e$ and $r_g$ by enforcing $\nabla_r^2\psi(r,z_0)=0$, ground potential in $r_g$ and a potential in $r_e$ computed from the flux condition Eq.~(\ref{Neumann}). This leads to the Dirichlet condition
\begin{equation}
    \psi(r,z_0) = \psi(r_e,z_0) \frac{\ln (r \big / r_g)}{\ln (r_e \big / r_g)}
    \label{Dirichlet}
\end{equation} 
which is consistent with the radial profile derived in Sec.~\ref{sec:III_C}. The rest of the domain's boundary conditions, as shown in figure~\ref{fig:FDM_BC}, are straightforward and imposed from symmetry or zero potential. 

To offer a more physical illustration of the effect of thermionic emission, we discuss here the results obtained for a set of dimensional geometric and plasma parameters (see Table~\ref{tab:table_FDM}) which correspond loosely to the conditions expected in a RF or Helicon laboratory plasma. 

\begin{table}[htbp]
\caption{\label{tab:table_FDM}Set of plasma and geometrical parameters used in numerical simulations.}
\begin{tabular}{lc}
\br
Neutral pressure $P$ [Pa]             & $0.15$\\ 
Plasma density $n$ [m$^{-3}$]             & $10^{18}$ \\ 
Electron temperature $T_e$ [eV] & $3$\\ 
Ion temperature $T_i$ [eV]     & $0.2$         \\ 
Magnetic field $B$ [mT]       & $140$                             \\
Cross section $\sigma_0$ [$10^{-19}$ m$^2$] 	     & $ 5 $ \\
Plasma column length $L$ [m]     & $2$                  \\
Plasma column radius $r_g$ [cm]	 & $20$ \\
Electrode radius $r_e$ [cm]        &  $2.5$\\
Electrode bias $\phi_e$ [V]          & $-250$ \\
Ion mass $m_i$ [amu] & $40$ \\
\hline
Normalized electrode potential $\psi_e$ &  $-83.3$ \\
Admissible plasma potential $\chi$ &  $1.5$ \\
Cold control parameter $|\psi_e|\chi$ &  $123$ \\
\br
\end{tabular}
\end{table}

\begin{figure}[htbp]
    \centering
    \includegraphics[scale = 1]{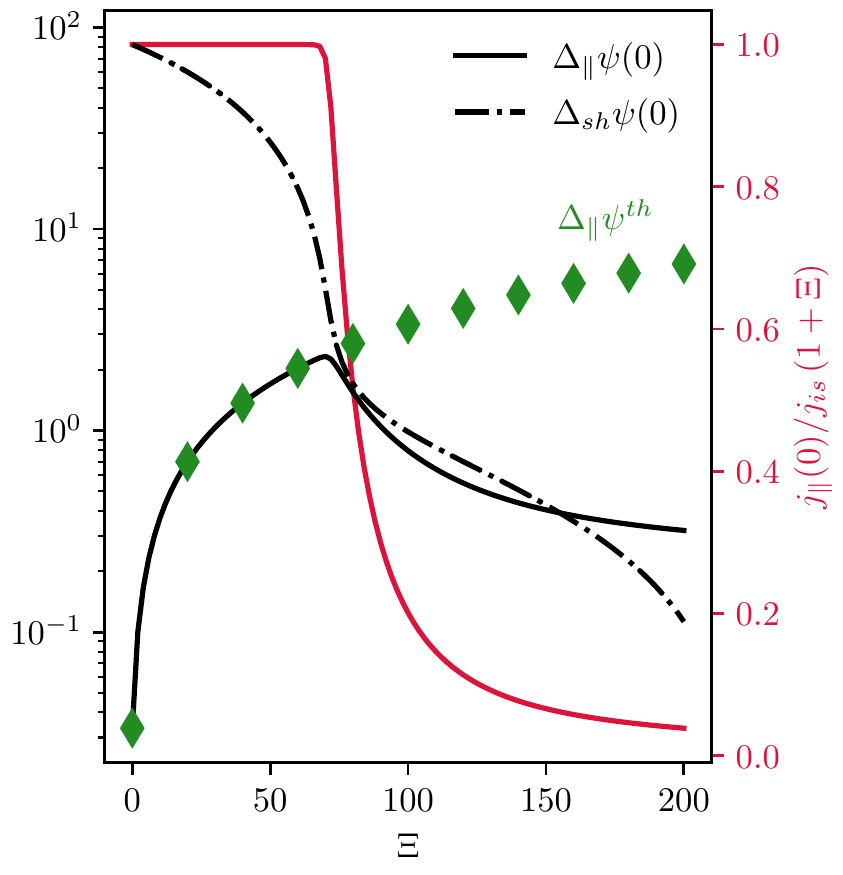}
    \caption{On-axis voltage drop across the sheath $\Delta_{sh} \psi(0)$ (black dashdot, left axis) and along field lines $\Delta_\parallel \psi(0)$ (black solid, left axis) and normalized current density (red solid, right axis) as a function of thermionic emission $\Xi$. The green diamond represents the analytical solution $\Delta_\parallel \psi^{sat}$ given in Eq.~(\ref{delta_para_psi_Ohm_Simplified}). Potentials are normalized by the electron temperature $T_e$ in eV, $\psi=\phi/T_e$.}
    \label{fig:Voltage_drop_Xi}
\end{figure}

Focusing first on the on-axis behavior, figure~\ref{fig:Voltage_drop_Xi} confirms as expected from the value of $|\psi_e|\chi\gg 1$ in Table~\ref{tab:table_FDM} that the regime is saturated for zero thermionic emission ($\Xi=0$). The current is indeed equal to its maximum value (in absolute value) while the voltage drop across the sheath $\Delta_{sh}\psi$ is significant. In these conditions the voltage drop along field lines $\Delta_{\parallel}\psi$ is very small. As the thermionic current is increased, that is as $\Xi$ increases, the voltage drop across the sheath $\Delta_{sh} \psi(0)$ decreases, as expected from the $|\psi_e|-(1+\Xi)\chi^{-1}$ scaling in the saturated regime. Meanwhile, the voltage drop along field lines increases, and numerical simulation results are observed to be in very good agreement with the theoretical prediction obtained in Eq.~(\ref{delta_para_psi_Ohm}). This trend continues up until $\Xi$ is large enough to lead to the transition from a saturated to a non-saturated regime on-axis. This transition is indicated by the sudden drop in normalized current in figure~\ref{fig:Voltage_drop_Xi}. Depending on plasma and geometrical parameters, the voltage drop along field line can already be larger than the voltage drop through the sheath when the regime transition takes place. In the non-saturated regime, both the voltage drop across the sheath and the voltage drop along field lines are observed to decrease with $\Xi$. 


\begin{figure*}[htbp]
    \centering
    \includegraphics[scale = 1]{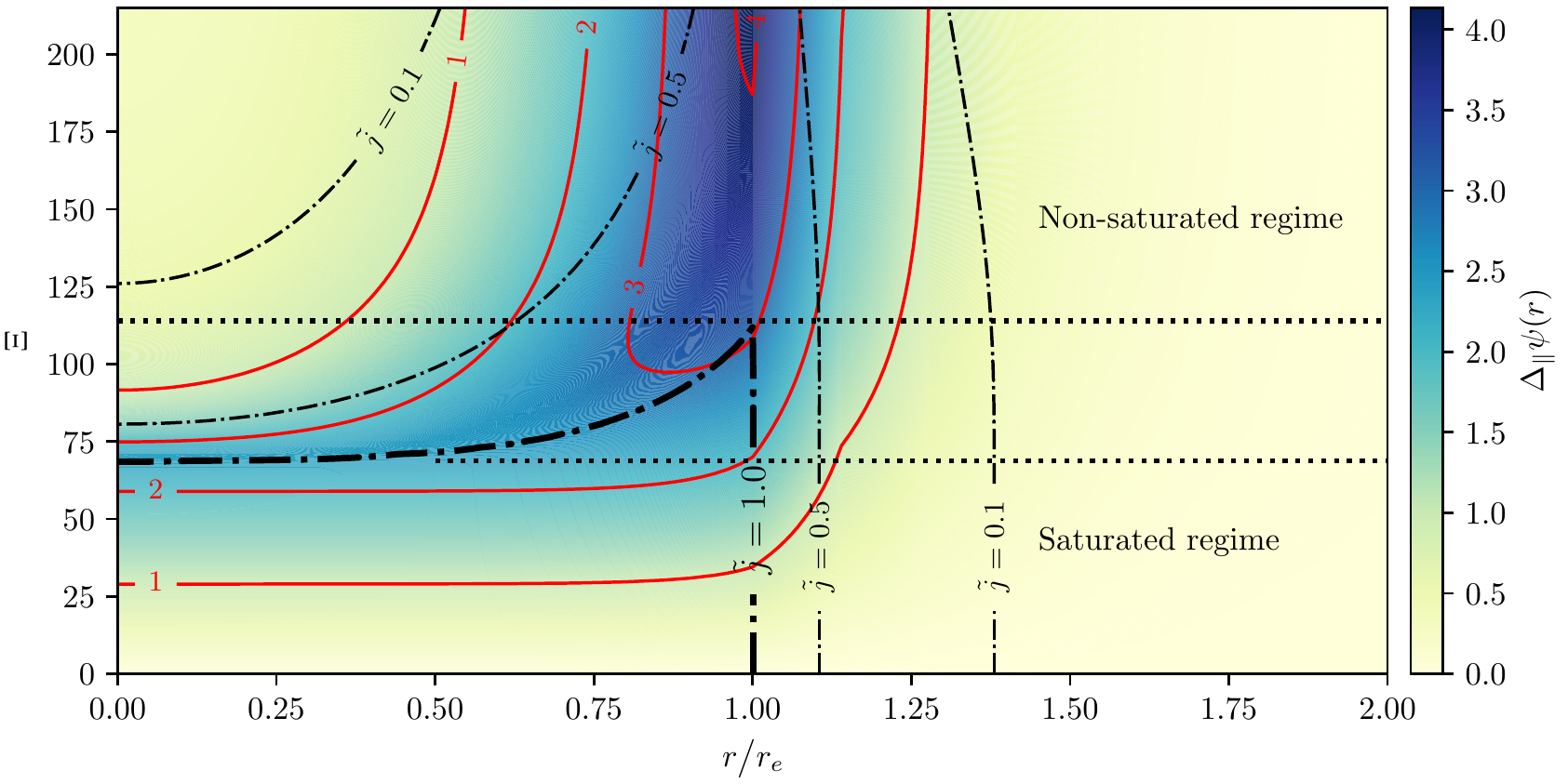}
    \caption{Colormap of the voltage drop along field lines $\Delta_\parallel \psi(r)$ as a function of the radius  $r$ and the thermionic emission $\Xi$. Red solid lines indicate iso-contours of $\Delta_\parallel \psi(r)$. Black dashdot lines indicate iso-contours of normalized current density $\tilde{j} = j_\parallel(r) \big / j_{max}$. The thick dashdot line represent the local regime transition, that is $\tilde{j}=1$. Black dotted lines indicate values of $\Xi$ for which transition for non-saturated to partially saturated and finally saturated regimes occur. Potentials are normalized by the electron temperature $T_e$ in eV, $\psi=\phi/T_e$.}
    \label{fig:test}
\end{figure*}

Additional insights can be gained by examining how this behavior depends on radius, as shown in figure~\ref{fig:test}. Starting again from zero emission ($\Xi=0$), one finds a uniform behavior across the radius, consistent with the fact that the regime is fully saturated. This is confirmed by the result that $\tilde{j}=j_\parallel(r)/ j_{max} = 1$ over the entire electrode. As thermionic emission is turned on and $\Xi$ increases, the voltage drop along field lines $\Delta_\parallel \psi$ is first observed to grow nearly uniformly in the electrode shadow ($r<r_e$). As $\Xi$ is increased further, the current iso-contours confirm that the transition from saturated to non-saturated regime occurs first on-axis, and progressively moves radially outward, as predicted in Section~\ref{sec:III_A}. This radial current density non-uniformity is accompanied by a similar variation in the voltage drop along field lines, consistent with Eq.~(\ref{delta_para_psi_Ohm}). For a given $\Xi$ the voltage drop along field lines $\Delta_{\parallel}\psi(r)$ hence grows with $r$ in this regime.

An interesting result revealed by figure~\ref{fig:test}, contrasting with on-axis results observed in figure~\ref{fig:Voltage_drop_Xi}, lies in the behavior in the non-saturated regime. Indeed, while both the voltage drop along field lines and the current density through the sheath on-axis were seen to drop rapidly with $\Xi$ once the non-saturated regime has been reached, one observes here that the voltage drop can actually grow with $\Xi$ in the outer region of the biased electrode, and that even in the non-saturated regime. The origin for this apparently counter-intuitive result is to be found in the fact the while the normalized current density $\tilde{j}$ falls off rapidly with $\Xi$ on-axis, the decrease is much slower near $r=r_e$. As a result the absolute parallel current density through the sheath can still locally grow with $\Xi$ near the outer radius of the electrode. As shown in \ref{app:A} this growth in parallel current can explain the observed growth of the voltage drop along field lines even in the non-saturated regime on the condition that the radial gradient length for the current is sufficiently large compared to the electrode radius, which is indeed what is observed in figure~\ref{fig:test}.

Returning to our motivation of assessing the potential drop along field lines, in particular in the case of thermionic emission, these results show that strong emission can lead to potential variations along field lines in the quasi-neutral plasma of the order of a few electron temperature. While moderate, these variations could be sufficient to affect perpendicular field homogeneity along field lines. This is particularly true when considering that, as discussed next, these variations may be even larger for a different set of operating conditions and geometric parameters. In addition, it is found here that the axial voltage drop can vary by a comparable amount perpendicular to field line across the radius of the biased electrode. This, combined with the possibility of a radially dependent voltage drop across the sheath, could lead to potential differences between field lines that intersect the same electrode, hindering the ability to control perpendicular electric fields.

\subsection{\label{sec:IV_C} Parametric dependencies}

While the results presented in figure~\ref{fig:Voltage_drop_Xi} and figure~\ref{fig:test} were obtained for specific plasma and geometric parameters, they can be used in combination with the analytical results obtained in Sec.~\ref{sec:IV_A} to offer a more general analysis. Indeed, although Eq.~(\ref{delta_para_psi_Ohm}) already provided insights into the effect of thermionic emission, a deeper understanding can be gained by developing the parametric dependencies of conductivities. 

Expressions for parallel and perpendicular conductivity in a magnetized plasma are in general non-trivial and require solving a full set of fluid equations\cite{song2001three,Rax2019,Kolmes2019} but one can often reasonably use simplified formulas within an appropriate operating parameter space. For instance, electron-neutral and ion-neutral collisions are found to be main contribution to respectively parallel and perpendicular conductivity across a range of low-temperature partly ionized and magnetized plasma laboratory experiments~\cite{Liziakin2021}. In this case the perpendicular conductivity reduces to the Pedersen conductivity
\begin{equation}
	\displaystyle \sigma_\perp = \frac{n m_i \nu_{in}}{{B_0}^2}
	\label{sigma_perp}
\end{equation}
while the parallel conductivity is
\begin{equation}
    \sigma_\parallel = \frac{e^2 n}{m_e \nu_{en}}
    \label{sigma_para}
\end{equation}
with $\nu_{en}$ and $\nu_{in}$ the electron-neutral and ion-neutral collision frequency, respectively. Here we simply take $\nu_{\alpha n} = \sigma_0 N_n\sqrt{T_\alpha/m_{\alpha}}$ with $\alpha=e,i$, $m_\alpha$ the particle mass, $\sigma_0$ a fiducial cross section and $N_n$ the neutral density, and ignore for simplicity the temperature dependence of $\sigma_0$. Plugging Eq.~(\ref{sigma_para}) into Eq.(\ref{delta_para_psi_Ohm}) yields
\begin{equation}
    \Delta_\parallel \psi^{sat}_{en} \propto \frac{\left ( 1 + \Xi \right ) L N_n}{ \sqrt{m_{i}}}.
    \label{delta_para_psi_Ohm_Simplified}
\end{equation}
Meanwhile, plugging Eq.~(\ref{sigma_perp}) into Eq.~(\ref{chi}) leads to 
\begin{equation}
    \displaystyle \chi_{in} \propto \frac{L\sqrt{T_eT_i} m_i N_n}{{r_{e}}^{2} \ln(r_g/r_e){B_0}^{2}}.
    \label{chi_puissance}
\end{equation}

Comparing Eq.~(\ref{delta_para_psi_Ohm_Simplified}) and Eq.~(\ref{chi_puissance}), one finds that in regimes where the simplified conductivities Eqs.~(\ref{sigma_perp}) and (\ref{sigma_para}) hold a decrease in the magnetic field intensity $B_0$ will increase the range of operating conditions leading to saturation. In particular, it will require a larger $\Xi$ to transition from saturated to non-saturated operation. Looking at figure~\ref{fig:test}, this means moving the transition region higher up. Meanwhile, since the voltage drop along field lines does not depend on $B_0$ but does depend on $\Xi$, the expansion of the saturated region means that higher voltage drop along field lines will be observed in this regime. From Eqs.~(\ref{sigma_perp}) and (\ref{sigma_para}), a similar growth is expected for heavier ion species since $\chi_{in}\propto m_i$ while $\Delta_\parallel \psi^{sat}_{en}\propto 1/\sqrt{m_i}$, or for a smaller electrode radius $r_e$.

The generic results Eq.~(\ref{chi}) and Eq.(\ref{delta_para_psi_Ohm}) can be similarly used to explore other regimes. For instance, if ones considers a denser plasma where parallel conductivity is now governed by Spitzer conductivity,
\begin{equation}
    \sigma_\parallel = \frac{e^2 n}{m_e \nu_{ei}}
    \label{sigma_spitzer}
\end{equation}
with $\nu_{ei}$ the electron-ion collision frequency then
\begin{equation}
    \Delta_\parallel \psi_{ei}^{sat} \propto \frac{\left ( 1 + \Xi \right ) L n }{\sqrt{m_{i}}{T_e}^{2}}.
\end{equation}
This last result shows that in these conditions an increase of the plasma density $n$ will lead to an increase in the voltage drop along field lines in the saturated regime. On the other hand the value of $\Xi$ for which transition occurs will decrease with $n$ since $\Xi\propto\sqrt{m_i/T_e}/n$ for $j_{eth}\gg j_{is}$. As a result the bias for which transition occurs will also decrease with $n$.

\section{\label{sec:V} Summary}

In this study the problem of plasma potential control from negatively biased emissive electrodes positioned at the axial ends of a magnetized plasma column is investigated through a combination of theory and numerical simulations.

By limiting the radial current flowing from the grounded vacuum vessel to the negatively biased electrode through the plasma, the ion sheath formed in front of the electrode is shown to control how much of the applied bias is passed on along field lines to the plasma potential. Besides geometrical parameters the minimum achievable plasma potential is controlled by the ion saturation current and the perpendicular plasma conductivity. When the applied bias amplitude is smaller than the voltage constructed from the plasma perpendicular resistance and the ion saturation current at the biased electrode, the plasma potential is found to be about the applied bias. In contrast, when the applied bias amplitude is larger than this voltage the plasma potential is no longer controlled by the applied bias, and the voltage difference is found across the sheath. These two regimes are referred to as non-saturated and saturated, respectively.

By increasing the current flowing through the plasma, thermionic emission from the biased electrode makes it possible to lower the minimum achievable plasma potential for a given set of plasma conditions and geometric parameters. This in turn allows accessing a broader range of plasma potential, or in other words improves plasma potential control through electrode biasing. On the other hand, thermionic emission from the biased electrode is shown to lead to a larger plasma potential variation along magnetic field lines in the quasi-neutral plasma. Scaling laws for this potential variation along field lines are derived analytically through the parametric dependencies of the plasma parallel and perpendicular conductivities, and verified against numerical simulations. Quantitatively, one finds plasma potential variations along field lines of a few electron temperature for typical RF or Helicon laboratory experiments under strong thermionic emission. 

Although such potential variations along field lines may not in themselves be a showstopper, they suggest that the possibility of using thermionic emission to minimize the voltage drop across the sheath and hence to maximize control over the plasma potential at the sheath edge should be consider carefully, and that beyond the practical limit resulting from the formation of a virtual cathode. This is particularly true in the non-saturated regime where a radially dependent axial voltage drop could lead to a rotation of the electric field vector.


\appendix
\section{\label{app:A} Analytical model for the voltage drop along field lines}

In the following we ignore thermionic emission for simplicity. Results accounting for thermionic emission can simply be obtained by substituting $j_{is}+j_{eth}$ in lieu of $j_{is}$ in the following results. 

Taylor expanding the first partial derivative of $\phi$ with respect to $z$ in $\left(r,z_0=-L/2\right)$ yields
\begin{align}
\frac{\partial \phi}{\partial z}(r,z) & = \frac{\partial \phi}{\partial z}\left(r,z_0\right)+\frac{\partial^2 \phi}{\partial z^2}\left(r,z_0\right)\left(z-z_0\right)\nonumber\\& \qquad+\frac{1}{2}\frac{\partial^3 \phi}{\partial z^3}\left(r,z_0\right)\left(z-z_0\right)^2+\mathcal{O}\left(z-z_0\right)^3.
\label{Eq:Taylor}
\end{align}
Additional terms in $\partial/\partial r$ expected for the expansion of a bi-variate function do not appear here since we only consider a change in position along $\mathbf{\hat{z}}$. From Ohm's law at the sheath edge the first term on the RHS of Eq.~(\ref{Eq:Taylor}) is
\begin{equation}
\frac{\partial \phi}{\partial z}\left(r,z_0\right) = -\frac{j_{sh,\parallel}}{\sigma_{\parallel}}.
\end{equation}
Then, from the anisotropic Laplace equation Eq.~(\ref{Eq:Laplace_anis}), the second and third partial derivative of $\phi$ with respect to $z$ write
\begin{equation}
\frac{\partial^2 \phi}{\partial z^2}(r,z) = -\frac{\sigma_{\perp}}{\sigma_{\parallel}}\frac{1}{r}\frac{\partial}{\partial r}\left(r\frac{\partial \phi}{\partial r}(r,z)\right)
\end{equation}
and 
\begin{align}
\frac{\partial^3 \phi}{\partial z^3}(r,z) & = -\frac{\partial }{\partial z}\left[\frac{\sigma_{\perp}}{\sigma_{\parallel}}\frac{1}{r}\frac{\partial}{\partial r}\left(r\frac{\partial \phi}{\partial r}(r,z)\right)\right]\nonumber\\
& = -\frac{\sigma_{\perp}}{\sigma_{\parallel}}\frac{1}{r}\frac{\partial}{\partial r}\left(r\frac{\partial}{\partial r}\frac{\partial \phi}{\partial z}(r,z)\right)
\nonumber\\
& = \frac{\sigma_{\perp}}{\sigma_{\parallel}}\frac{1}{r}\frac{\partial}{\partial r}\left(r\frac{\partial}{\partial r}\frac{j_{\parallel}}{\sigma_{\parallel}}(r,z)\right).
\label{Eq:third_derivative}
\end{align}

Writing $\bar{r} = r/r_e$ and $\bar{z} = z/L$, Eq.~(\ref{Eq:third_derivative}) rewrites
\begin{equation}
    \frac{\partial^3 \phi}{\partial \bar{z}^3}(\bar{r},\bar{z}) = \tau^2 L\left(\frac{r_g}{r_e}\right)^2 \frac{1}{\bar{r}}\frac{\partial}{\partial \bar{r}}\left(\bar{r}\frac{\partial}{\partial \bar{r}}\frac{j_{\parallel}}{\sigma_{\parallel}}(\bar{r},\bar{z})\right)
\end{equation}
with 
\begin{equation}
    \tau=\frac{L}{r_g}\sqrt{\frac{\sigma_{\perp}}{\sigma_{\parallel}}}
\end{equation}
as previously defined in Ref.~\cite{Gueroult2019a}. This qualitatively shows that the contribution of the third partial derivative in Eq.~(\ref{Eq:Taylor}) scales as $(\tau/L_{\nabla_r} j_{\parallel})^2$ where ${L_{\nabla_r} j_{\parallel}}$ characterizes the radial gradient length of the parallel current density in units of $r_e$.

Similarly, higher order derivatives can be written as
\begin{equation}
    \frac{\partial^n \phi}{\partial \bar{z}^n}(\bar{r},\bar{z}) = \tau^2L\left(\frac{r_g}{r_e}\right)^2\frac{1}{\bar{r}}\frac{\partial}{\partial \bar{r}}\left(\bar{r}\frac{\partial}{\partial \bar{r}}\left[\frac{\partial^{n-3}}{\partial \bar{z}^{n-3}}\frac{j_{\parallel}}{\sigma_{\parallel}}(\bar{r},\bar{z})\right]\right).
\end{equation}
A similar analysis as that done for the third derivative can thus be made here, with the difference that the gradient length is now taken on the $(n-3)^{th}$ partial derivative of $j_{\parallel}$ with respect to $\bar{z}$. Because $j_{\parallel}$ is expected to vary smoothly along field lines from its value at the electrode to $0$ for $z=0$ (by symmetry), the contribution of these extra terms is expected to decrease rapidly with $n$, and in any case to be dominated by third order contribution. 

In the saturated regime the parallel current density is constant along the entire electrode, $j_{sh,\parallel} = -j_{is}$, and thus $j_{\parallel}(r\leq r_e,z_0)$ does not depend on $r$ (\emph{i.~e.} $L_{\nabla_r}j_{\parallel}\rightarrow\infty$). Eq.~(\ref{Eq:third_derivative}) then implies that the last term in Eq.~(\ref{Eq:Taylor}) simply cancels out in this case. In addition, 
\begin{equation}
\frac{\partial \phi}{\partial r}\left(r,z_0\right) = \frac{j_{is}r}{L\sigma_{\perp}}
\end{equation}  
so that 
\begin{equation}
\frac{1}{r}\frac{\partial}{\partial r}\left(r\frac{\partial \phi}{\partial r}\right)\left(r,z_0\right) = 2\frac{j_{is}}{L\sigma_{\perp}}.
\end{equation}
Putting these pieces together Eq.~(\ref{Eq:Taylor}) finally leads to
\begin{equation}
\frac{\partial \phi}{\partial z}(r,z) = \frac{j_{is}}{\sigma_{\parallel}}\frac{2z}{L}\left[1+ \mathcal{O}\left(\tau^2\right)\right]
\end{equation}
where the $\mathcal{O}\left(\tau^2\right)$ term comes from higher order terms involving $\partial j_{\parallel}/\partial z$. Integration between $z_0$ and $0$ with the boundary conditions $\phi(r,z_0) = \phi_{sh}(r)$ and $\phi(r,0) = \phi_{mid}(r)$ yields to lowest order in $\tau$
\begin{equation}
    \Delta_{\parallel}\phi^{sat}(r) = \frac{L}{4\sigma_{\parallel}}j_{is}.
    \label{Eq:appendix_voltage_drop}
\end{equation}

In the non-saturated regime, the contribution of the third derivative is no longer zero but instead scales as $(\tau/L_{\nabla_r} j_{\parallel})^2$. This suggests that an analog to Eq.~(\ref{Eq:appendix_voltage_drop})
\begin{equation}
    \Delta_{\parallel}\phi(r) = -\frac{L}{4\sigma_{\parallel}}j_{sh,\parallel}(r).
\end{equation}
roughly holds as long as 
\begin{equation}
    j_{sh,\parallel}(r,z_0)\Big / \left[r_e\frac{\partial j_{sh,\parallel}(r,z_0)}{\partial r}\right]\gg\tau.
\end{equation}

\section*{References}

\providecommand{\noopsort}[1]{}\providecommand{\singleletter}[1]{#1}%
\providecommand{\newblock}{}

\end{document}